\documentclass[12pt,epsfig]{article}
\usepackage{graphicx,amsmath,amssymb}

\newlength{\dinwidth}
\newlength{\dinmargin}
\def\lapproxeq{\lower .7ex\hbox{$\;\stackrel{\textstyle
<}{\sim}\;$}}
\def\gapproxeq{\lower .7ex\hbox{$\;\stackrel{\textstyle
>}{\sim}\;$}}
\def\gtrsim{\lower .7ex\hbox{$\;\stackrel{\textstyle
>}{\sim}\;$}}
\def\lesim{\lower .7ex\hbox{$\;\stackrel{\textstyle
<}{\sim}\;$}}

\def\be{\begin{equation}}
\def\ee{\end{equation}}
\def\bea{\begin{eqnarray}}
\def\eea{\end{eqnarray}}
\def\funp{{I\!\!P}}

\def\bb{b\bar{b}}

\def\ra{ \rightarrow }

\begin{document}
\begin{flushright}
IPPP/07/04 \\
DCPT/07/08 \\
21st February 2007 \\

\end{flushright}

\vspace*{0.5cm}

\begin{center}
{\Large \bf Insight into Double-Pomeron-Exchange Higgs production and backgrounds}

\vspace*{1cm}
\textsc{V.A.~Khoze$^{a,b}$, A.D. Martin$^a$ and M.G. Ryskin$^{a,b}$} \\

\vspace*{0.5cm}
$^a$ Department of Physics and Institute for
Particle Physics Phenomenology, \\
University of Durham, DH1 3LE, UK \\
$^b$ Petersburg Nuclear Physics Institute, Gatchina,
St.~Petersburg, 188300, Russia \\

\end{center}

\vspace*{0.5cm}

\begin{abstract}
We quantify the central inclusive background contributions to exclusive Higgs production at the LHC arising from double-Pomeron-exchange processes. We consider the $H \to \bb$ signal. We study processes mediated by the fusion of two `hard' Pomerons, and also by the fusion of `soft' Pomerons. The latter background is found to be very small, and the former is found to be less than the exclusive signal.
\end{abstract}

The experimental observation of exclusive Higgs production, $pp \to p+H+p$, has some attractive features. In particular, it is possible (i) to measure the mass of the Higgs accurately by tagging the outgoing protons and measuring the missing mass, (ii) to study the quantum numbers of the centrally produced Higgs system, and (iii) to observe Higgs production in an environment with a relatively low background. First there are no soft secondaries generated by the underlying event, and, second, due to a ``$J_z=0$'' selection rule \cite{KMRmm}, leading-order QCD $\bb$ production is suppressed by a factor $(m_b/E_T)^2$, where $E_T$ is the transverse energy of the $b$ jets. Therefore for a low mass Higgs, $M_H \lapproxeq 150$ GeV, there is a chance to observe the main $\bb$ decay mode \cite{KMRpros,DKMOR,KMRrev}, and to directly measure the $H \to \bb$ Yukawa coupling constant. 

\begin{figure}
\begin{center}
\includegraphics[height=5cm]{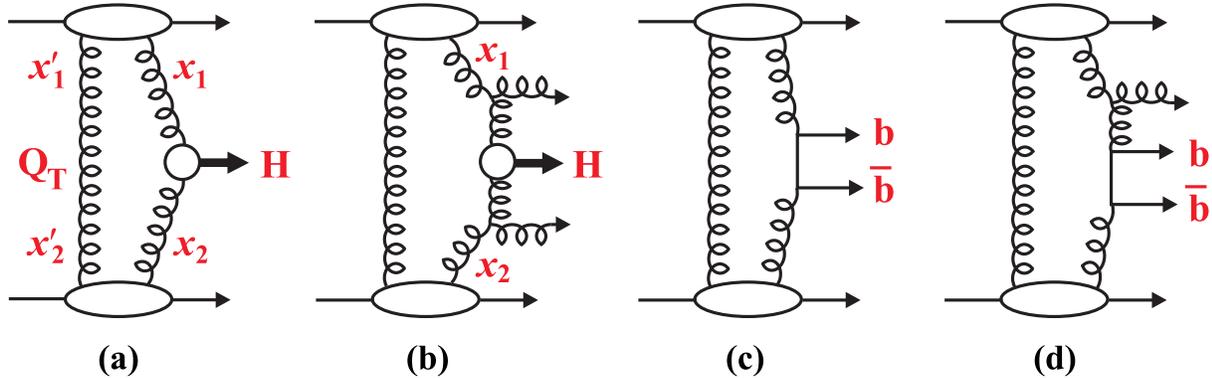}
\caption{(a) Exclusive Higgs production by the fusion of two hard Pomerons; (b) Higgs production, via hard Pomerons, but accompanied by the emission of two undetected gluons; (c,d) background QCD $\bb$ production processes. For (b,c,d) we account for the full set of Feynman diagrams at this order and, moreover, for (b,d) allow for additional soft gluon emission.}
\label{fig:1}
\end{center}
\end{figure}

\begin{figure}
\begin{center}
\includegraphics[height=8cm]{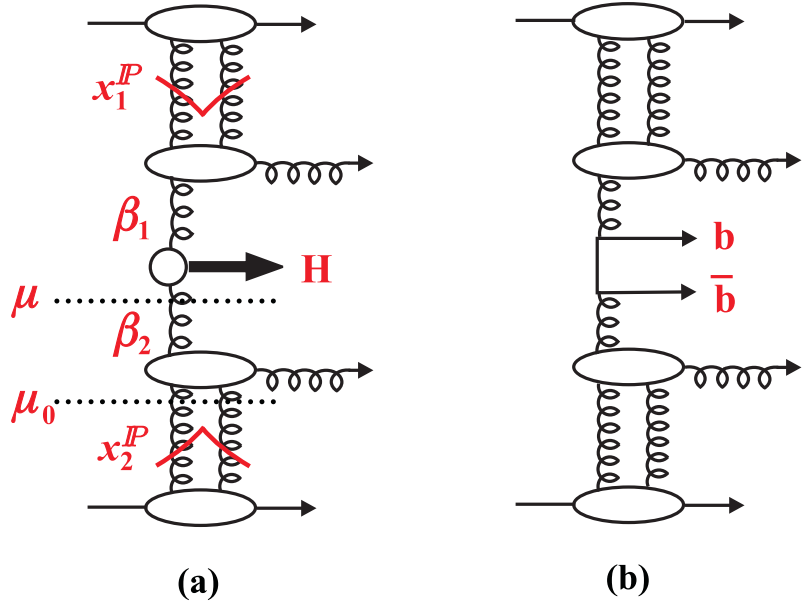}
\caption{(a) Higgs production by the fusion of two soft Pomerons; (b)  background $\bb$ production via soft Pomerons.}
\label{fig:2}
\end{center}
\end{figure}

The process $pp \to p+H+p$, where the + signs denote large rapidity gaps, may be described by the fusion of two Pomerons into the Higgs boson. However we must distinguish between the contributions from a small size Pomeron (which can be described by perturbative QCD) and a relatively large size Pomeron (which must be described phenomenologically). Here, we call these {\it hard} and {\it soft} Pomerons respectively. Higgs production (and the background processes) resulting from the fusion of two hard Pomerons is shown diagrammatically in Fig.~\ref{fig:1}, and those obtained from soft Pomeron fusion are shown in Fig.~\ref{fig:2}. {\it Only} Fig.~\ref{fig:1}(a) corresponds to the {\it exclusive} Higgs signal. Its cross section can be calculated by perturbative QCD, upto an overall survival factor, ${\hat S}^2$, for the large rapidity gaps; for a recent review and references see \cite{uri}.  We will briefly recall the main steps in the calculation in a moment. The remaining diagrams in Figs.~\ref{fig:1} and \ref{fig:2} correspond to processes which may contaminate the exclusive signal.  It is the purpose of this note to quantify these background contributions.


The soft Pomeron contributions of Fig.~\ref{fig:2} will be evaluated using an Ingelman-Schlein-like approach \cite{IS}, in which, at some rather low scale $\mu_0$, the incoming Pomerons may be treated as hadrons whose wave functions contain gluons, followed by the use of the usual collinear factorization formula. Thus the central system is produced by the fusion of gluons from two different soft Pomerons. Formally, these are {\it not} exclusive processes. Besides the Higgs boson (or $\bb$ pair), we have at least two secondary gluons --- the remnants of the Pomerons. However, if these gluons have low $k_T$ they may not be observed experimentally. Thus the soft double-Pomeron-exchange processes of Fig.~\ref{fig:2} may fake exclusive signals\footnote{Note that, from a formal point of view, the soft Pomeron amplitudes of Fig.~\ref{fig:2}(a,b) are suppressed by an additional power of $\alpha_S$ in comparison with the corresponding amplitudes arising from hard Pomerons \cite{KMRpros,KMRmyth}. Therefore the contributions of Fig.~\ref{fig:2}(a,b) may be important only at low scales $\mu_0$ associated with the Pomeron-parton splitting vertex, where $\alpha_S$ is not too small.}. A similar problem may occur for the hard Pomerons, where undetected gluons may be emitted from the active $t$-channel gluons, labelled $x_1$ and $x_2$ in Fig.~\ref{fig:1}(b)\footnote{Emissions from the screening gluon $Q_T$ are very small \cite{KMRmyth,KRS}.}; for perturbative QCD $\bb$ production of Fig.~\ref{fig:1}(d), it is possible to emit only one extra gluon, since its colour can be compensated by the colour of the $\bb$ pair.

As mentioned above, the purpose of the present note is to evaluate the size of the backgrounds arising from the types of process shown in Figs.~\ref{fig:1} and \ref{fig:2}, to the exclusive process of Fig.~\ref{fig:1}(a). That is the background due to additional gluons which escape experimental detection. We consider both Higgs production and the QCD production of a $\bb$ pair, which is one of the main sources of background to the exclusive $H \to \bb$ signal. Unlike the irreducible background of Fig.~\ref{fig:1}(c), the events with additional gluon emission can, in principle, be separated from exclusive production. However the acceptance and efficiency of the detectors is less than 100$\%$, so parts of these types of contribution must be considered as background to the exclusive Higgs signal. To suppress this background it is natural to use the matching condition \cite{DKMOR},
\be
M_{\rm missing}~=~M_{\rm central},
\label{eq:mm}
\ee
between the mass of the centrally produced system (i) as measured by the {\it missing} mass to the tagged very forward outgoing protons, and (ii) as measured by the mass of the $\bb$ dijet system observed in the {\it central} detector. Unfortunately, the reconstruction of the dijet mass is not very precise. Here we shall allow for a mismatch $\Delta M/M$ of 20$\%$, which is twice the expected mass resolution of the central detector. Nevertheless, the matching condition, even with this $\Delta M$ tolerance, puts a strong constraint on the momentum fractions, $z_i$, carried by the undetected gluon emissions. For example, for Fig.~\ref{fig:1}(b), the requirement that the observed $\bb$ dijet mass should not be much smaller than the mass of the central system as measured with good accuracy by the missing mass to the tagged protons, leads to the constraint
\be
\frac{M_{\rm dijet}}{M_{\rm missing}}~=~\sqrt{(1-z_1)(1-z_2)}~>~\left(1-\frac{\Delta M}{M}\right).
\label{eq:MMcons}
\ee
So the fractions $z_i$ are required to be small, that is the Pomeron momentum fractions, $\beta_i = (1-z_i)$, carried by the active gluons should be close to 1.

It is convenient to first compute the background arising from the soft Pomeron contributions of Fig.~\ref{fig:2}. Often the name Double-Pomeron-Exchange (DPE) is reserved for these types of process. The cross sections can be readily calculated in terms of the usual collinear factorization formula. For the Higgs signal of Fig.~\ref{fig:2}(a), we have
\be
\sigma~=~{\hat S}^2_{\rm DPE}\int dx_1^{\funp} dx_2^\funp d\beta_1 d\beta_2~g^D(\beta_1,x_1^\funp,\mu)g^D(\beta_2,x_2^\funp,\mu)\hat{\sigma}(gg \to H) ~\theta\left[\sqrt{\beta_1\beta_2}-\left(1-\frac{\Delta M}{M}\right)\right],
\label{eq:soft}
\ee
where $g^D$ are the diffractive gluon distributions at scale $\mu \sim M_H/2$, and the $\theta$-function reflects the fact that the observed dijet mass should not be much smaller than the measured missing mass, see (\ref{eq:MMcons}). The HERA data for inclusive deep inelastic diffractive scattering are not sufficient to fix the behaviour of the gluon distribution, $g^D$, at large $\beta$. The H1 collaboration present \cite{H1} two preliminary fits (A and B) to their recent inclusive diffractive data. In fit A the gluon distribution, at scales $\mu^2 \sim 10~{\rm GeV}^2$, is almost constant for $\beta \gapproxeq 0.5$, whereas in fit B it decreases rapidly with increasing $\beta$. However only fit B is consistent with the data on diffractive charm and diffractive dijet production. The MRW analysis \cite{MRW}, which allows for the perturbative QCD contribution, gives gluons which have analogous large $\beta$ behaviour to those of fit B. Here we will use the diffractive gluon distribution, $g^D$, found in the recent MRW analysis.  Since the diffractive gluons, $g_D$, vanish\footnote{Note that the large scale $\mu \sim M_H/2$ the QCD radiation leads to an additional power suppression for values of $\beta$ close to 1.} as $\beta \to 1$, the cross section obtained from (\ref{eq:soft}) turns out to be small; and much less than the previous evaluations based on the POMWIG Monte Carlo \cite{pomwig} using old diffractive gluon distributions which were approximately flat for $\beta \gapproxeq 0.5$, analogous to those of the H1 fit A. The factor $S^2_{\rm DPE}$ is the probability that the rapidity gaps associated with the soft Pomeron exchanges survive population from secondaries from the underlying event. The value that we use for $S^2_{\rm DPE}$ is given in the footnote of the paragraph containing (\ref{eq:M}).

The QCD $\bb$ production arising from soft gluons,  Fig.~\ref{fig:2}(b), is calculated using the same expression, (\ref{eq:soft}), but with the replacement of the cross section of the signal subprocess, $\hat{\sigma}(gg \to gHg \to g\bb g)$, by that for the QCD subprocess, $\hat{\sigma}(gg \to g\bb g)$.

Experimentally it is not possible, and not needed, to access these DPE contributions over the whole phase space. Here we present results for $d\sigma/dy$ at $y=0$ for the contribution collected in a missing mass window with $\Delta M_{\rm missing}=4$ GeV, that looks realistic bearing in mind the mass resolution expected from the roman pots which have been proposed to tag the forward protons \cite{FP420}.  Thus in the cross section for the process shown in Fig.~\ref{fig:2}(a) the integrals over $x_i^{\funp}$ in (\ref{eq:soft}) are replaced by the factor $\Delta M_{\rm missing}/\Delta M$, since in the limit of a small Higgs width, the $\hat{\sigma}(gg \to H)$ cross section contains a factor $\delta(M^2_{\bb}-M^2_H)$ spread over the $\Delta M$ interval.  (Recall that $\Delta M$ is controlled by the mass resolution of the $\bb$ dijet system, which is expected to be much larger than $\Delta M_{\rm missing}$.) On the other hand the QCD $\bb$ production of Fig.~\ref{fig:2}(b) has no corresponding $\delta$-function. Thus the $dx_i^{\funp}$ integrals are replaced by the usual logarithmic $dM^2/M^2$ phase space integral, leading to a factor $2\Delta M_{\rm missing}/M$ in the differential cross section $d\sigma/dy$.

The predictions for the processes of Fig.~\ref{fig:2} are shown in the Table. Since the diffractive gluon density $g^D$ decreases rapidly as $\beta$ increases, the $gHg$ background of Fig.~\ref{fig:2}(a) is negligible and the $g\bb g$ contribution of Fig.~\ref{fig:2}(b) is very small. 

\begin{table}[htb]
\begin{center}
\begin{tabular}{|l|c|c|c|}\hline

process  &  diagram  &  cross section  &   \\ \hline

$\sigma_{\rm excl}(H \to \bb)$  &   Fig.~\ref{fig:1}(a) & 150  &  exclusive signal  \\
$\sigma(gHg)$  &  Fig.~\ref{fig:1}(b)  &  20  &   \\

$\sigma^{\rm QHNC}(\bb: {\rm LO})$  &  Fig.~\ref{fig:1}(c)  & 70  & irreducible background \\
$\sigma^{\rm QHNC}(\bb g)$  &  Fig.~\ref{fig:1}(d)  &  5.2  &  $\lambda$ not conserved \\ 
$\sigma^{\rm QHC}(\bb g)$  &  Fig.~\ref{fig:1}(d)  &  0.6  &  negligible; $\lambda$ conserved \\ \hline

$\sigma^{\rm DPE}(gHg)$  &  Fig.~\ref{fig:2}(a)  &  0.14  &  negligible \\ 

$\sigma^{\rm DPE}(g\bb g)$  & Fig.~\ref{fig:2}(b)  &  9  &  small \\ \hline
\end{tabular}
\end{center}
\caption{The cross sections $d\sigma/dy|_{y=0}$ (in units of $10^{-3}$ fb) of the hard Pomeron processes shown in Fig.~\ref{fig:1} and the soft Pomeron processes of Fig.~\ref{fig:2}. In each case the $H \to \bb$ branching ratio has been included and a polar-angle cut $60^\circ<\theta(b)<120^\circ$ in the Higgs rest frame has been applied to the $b$ jet, that is the jet rapidity separation $|\eta_1-\eta_2|<1.1$.  We have taken $M_H=120$ GeV, and assumed that the mass resolutions of the central detector and roman pots are such that they correspond to mass windows $\Delta M_{\rm dijet}/M_{\bb}=20\%$ and $\Delta M_{\rm missing}=$ 4 GeV, respectively. $\lambda$ is the helicity along the $b$ quark line. For the processes of Fig.~\ref{fig:1}(b,d), we allow for the emission of any number of gluons with transverse momentum $k_T<k_{T,{\rm max}}=5$ GeV.}
\end{table}

We now turn to the processes mediated by hard Pomeron exchange.
The background contributions of Fig.~\ref{fig:1}(b,d) are calculated in a similiar way to that which was used to determine the exclusive Higgs signal of Fig.~\ref{fig:1}(a). So first we recall the latter calculation, see, for example, \cite{KMR}. 
The exclusive Higgs cross section is written as the convolution \cite{KMRpros} $\sigma_{\rm excl}={\cal L}{\hat \sigma}$ where the effective luminosity
\begin{equation}
{\cal L} ~\simeq ~\frac{{\hat S}^2}{b^2} \left|\frac{\pi}{8} \int\frac{dQ^2_T}{Q^4_T}\: f_g(x_1, x_1', Q_T^2, \mu^2)f_g(x_2,x_2',Q_T^2,\mu^2)~ \right| ^2~,
\label{eq:M}
\end{equation}
and the $gg^{\funp\funp} \to H$ subprocess cross section is
\be
{\hat \sigma}(M^2)~=~2\pi^2\frac{\Gamma(H \to gg)}{M_H^3}\delta\left(1-\frac{M^2}{M_H^2}\right).
\label{eq:higgs}
\ee
The $\funp\funp$ superscript on $gg$ is to emphasize that each gluon comes from a colour-singlet $t$-channel state. The first factor in (\ref{eq:M}), ${\hat S}^2 $, is the probablity that the rapidity gaps survive against population by secondary hadrons from the underlying event, that is hadrons originating from {\it soft} rescattering; and $b$ is the $t$-slope which describes the forward proton $p_T^2$ distribution. The value of ${\hat S}^2 $ was calculated using a model \cite{KMRsoft} which describes soft hadronic data in the CERN-ISR to Tevatron energy range. Since at large impact parameters the opacity of the proton decreases, the rapidity gaps of processes with a larger $t$-slope have a greater probability to survive. Interestingly, in \cite{KMRS} it was demonstrated that the ratio ${\hat S}^2/b^2$ is almost constant for relevant interval of values of the slope\footnote{For the soft $\funp\funp$ production processes of Fig.~\ref{fig:2} the corresponding slope $b$ is larger than that for the exclusive process of Fig.~\ref{fig:1}(a), due to the slope $\alpha'$ of the Regge trajectory of the soft Pomeron. As a result we find, using the observed slope of leading protons from diffractive deep inelastic scattering at HERA \cite{H1}, that ${\hat S}_{\rm DPE}^2 =2.3{\hat S}^2 ~=0.06$.} $b$.  At the LHC the value is found to be ${\hat S}^2/b^2 \simeq 0.0015~{\rm GeV}^4$ for $pp\ra p+H+p$.
The remaining factor, $|...|^2$, however, may
be calculated using perturbative QCD techniques, since the dominant contribution to the integral comes from the region $\Lambda_{\rm QCD}^2\ll Q_T^2\ll M_H^2$.
The probability amplitudes, $f_g$, to find the appropriate pairs of
$t$-channel gluons are given by the skewed
  unintegrated gluon densities at a {\it hard} scale $\mu \sim M_H/2$.

Since the momentum fraction $x'$ transfered through the
screening gluon $Q$ is much smaller than that ($x$) transfered through
the active gluons $(x'\sim Q_T/\sqrt s\ll x\sim M_H/\sqrt s\ll 1)$, it
is possible to express $f_g(x,x',Q_T^2,\mu^2)$
in terms of the conventional integrated density
$g(x)$ \cite{MR}. A simplified form of this relation is \cite{KMR}
\begin{equation}
\label{eq:a61}
  f_g (x, x^\prime, Q_T^2, \mu^2) \; = \; R_g \:
\frac{\partial}{\partial \ln Q_T^2}\left [ \sqrt{T_g (Q_T, \mu)} \: xg
  (x, Q_T^2) \right ],
\end{equation}
which holds to 10--20\%
accuracy.
The factor $R_g$ accounts for
the single $\log Q^2$ skewed effect.  It is found to
be about 1.2 at the energy of the LHC.

Note that the $f_g$'s embody a Sudakov suppression
factor $T_g$, which ensures that the gluon does not radiate in the
evolution from $Q_T$ up to the hard scale $\mu \sim M_H/2$, and so
preserves the rapidity gaps. The Sudakov factor is \cite{WMR,KMR}
\begin{equation}
\label{eq:a71}
  T_g (Q_T, \mu)=\exp \left (-\int_{Q_T^2}^{\mu^2}
  \frac{\alpha_S (k_T^2)}{2 \pi}\frac{dk_T^2}{k_T^2} \left[
  \int_\Delta^{1-\Delta}zP_{gg} (z)dz
\ + \ \int_0^1 \sum_q\
  P_{qg} (z)dz\right]\right),
\end{equation}
with $\Delta = k_T/(\mu + k_T)$.  The square root arises in
(\ref{eq:a61}) because the (survival) probability not to emit any
additional gluons  is only relevant to
the hard (active) gluon.  It is the presence of this Sudakov factor
which makes the integration in (\ref{eq:M}) infrared stable, and
perturbative QCD applicable.  

The resulting cross section for the exclusive production of a Higgs boson of mass $M_H=120$ GeV is 2.6 fb at the LHC. Allowing for the polar-angle acceptance cut (given in the Table caption) and the $H \to \bb$ branching ratio reduces this by a factor of about 3. The Table lists the corresponding cross section $d\sigma/dy$ at $y=0$.

Now let us allow for extra gluon radiation, which we assume goes undetected if it has transverse momentum $k_T<k_{T, \rm max}$. The diagrams of Fig.~\ref{fig:1}(b,d) show the simplest background contributions of this type, with, respectively, two gluons and a single undetected gluon. First we evaluate the process of Fig.~\ref{fig:1}(b). We might anticipate that this background will cancel the part of the exponent in the Sudakov factor of (\ref{eq:a71}) which is coming from the phase space which may be occupied by the emitted gluons (as would happen in the analogous case of photon emission in QED). However, in QCD, the colour correlations lead to an additional suppression of the real coloured gluon emission, so that this cancellation is incomplete. The colour correlations take the following form \cite{Cinel}. The active gluons, $g(x_1),~g(x_2)$ are correlated in colour with the screening gluon $Q_T$ in the amplitude $M$, and not with the active gluons $g'$ in the complex conjugate amplitude $M^*$. So the uncorrelated active $t$-channel gluons $g$ and $g'$, in $M$ and $M^*$ respectively, form the colour multiplets $i=1,8,10,27$ with the probabilities given by the statistical weights
\be
c_i~=~~\frac{1}{64},~~\frac{8+8}{64},~~\frac{10+10}{64},~~\frac{27}{64}~.
\ee
The colour factors for real emission of gluons in each of these multiplets are
\be
\lambda_i~=~~3,~~3/2,~~0,~~-1,
\ee
respectively. These real emissions lead to an exponent analogous to that in the Sudakov form factor of the virtual emissions, $T_g$ of (\ref{eq:a71}), multiplied by the corresponding colour factor $\lambda_i/N_c$.
Then taking each exponent with its weight $c_i$, we obtain
\be
T_g^{(\rm real)}~=~\sum_i c_i~\exp
  \left ( \frac{\lambda_i}{N_c}\int_{Q_T^2}^{k^2_{T,{\rm max}}}
  \frac{\alpha_S (k_T^2)}{2 \pi}\frac{dk_T^2}{k_T^2}
  \int_\Delta^x P_{gg} (z)dz \right)~ ,
\label{eq:Treal}
\ee
where $x=2\Delta M/M$, which guarantees that the emitted gluon does not violate the mass matching condition (\ref{eq:MMcons}). Note that for the colour singlet, $\lambda_1/N_c=1$,  we do get exact cancellation between the real and virtual terms (as would happen in QED), but this only happens with a probability of 1/64. Summing over $n$ soft emissions we have
\be
\sum_{n=0}^\infty~\sigma(H+ng)~=~\sigma_{\rm exclusive}~T_g^{(\rm real)}.
\label{eq:ng0}
\ee
For example, for the emission of only one gluon, we expand the exponent and note $\sum c_i\lambda_i=0$, which reflects the impossiblity of emitting only one gluon when producing a (colourless) Higgs boson. The right-hand-side of (\ref{eq:ng0}) has a symbolic form. Actually $T_g^{(\rm real)}$ must be included inside the $Q_T$ integral in (\ref{eq:M}). From (\ref{eq:ng0}), the
background contribution of Fig.~\ref{fig:1}(b) is given by
\be
\sum_{n=1}^\infty~\sigma(H+ng)~=~\sigma_{\rm exclusive}~(T_g^{(\rm real)}-1)~\frac{\Delta M_{\rm missing}}{\Delta M},
\label{eq:ng}
\ee
where now we include the mass resolution factors. Note that with decreasing $\Delta M$ the upper limit $x$ in the last integral in (\ref{eq:Treal}) decreases and becomes close to the lower limit $\Delta$. In terms of (\ref{eq:ng}), it means that $ T_g^{(\rm real)} \to 1$ and that the probability of extra gluon emission vanishes. The final result for the cross section of the process of Fig.~\ref{fig:1}(b), which we label $\sigma (gHg)$, is shown in the Table.

To calculate the background contribution of Fig.~\ref{fig:1}(d) we need only replace the hard subprocess Higgs production cross section, (\ref{eq:higgs}), by the cross section for the subprocess $gg^{\funp\funp} \to \bb g$. At collinear leading-order, the contribution (QHC) to this cross section, assuming helicity is conserved along the $b$-quark line, is given by \cite{KRS} 
\be
d\hat{\sigma}^{\rm QHC} ~=~ \frac{9}{64}~\frac{z^4}{(1-z)^3}~\frac{dz}{z}~\frac{dk_T^2}{k_T^2}~\frac{\alpha_S^3}{M^2_{\rm central}}\left(\frac{{\rm cos}^2\theta(1+{\rm cos}^2\theta)}{1-{\rm cos}^2\theta}\right)d{\rm cos}\theta,
\label{eq:QHC}
\ee
where $z$  and $k_T$ are the momentum fraction and transverse momentum of the emitted gluon; the factor 9/64 embraces the colour factor, and $\theta$ is the polar angle of the $b$-jet in the $\bb$ rest frame. The equation accounts for the emission of the $g$-jet in both the $x_1$ and $x_2$ directions. Again, we allow for the emission of additional soft gluons via $T_g^{(\rm real)}$, as in (\ref{eq:ng0}). Since now we no longer have the delta function $\delta(1-M^2/M^2_H)$ of (\ref{eq:higgs}), we instead must include $dM^2/M^2$, as we have done for Fig.~\ref{fig:2}(b). That is, the factor $\Delta M_{\rm missing}/\Delta M$ of (\ref{eq:ng}) is replaced by $2\Delta M_{\rm missing}/M$.
An important feature is that, due to the $J_z=0$ selection rule, the probability to emit soft gluons contains a suppression factor $z^4$ in the massless quark limit $m_b \ll M_H$, see (\ref{eq:QHC}) \cite{KRS,BKSO}. On the other hand, to satisfy the mass matching condition, (\ref{eq:MMcons}), we must have small $z$. It is therefore not surprising that we find that the QHC contribution has a negligibly small cross section.

The cross section (\ref{eq:QHC}) only accounts for the ``QHC'' contribution, where helicity is conserved along the $b$ quark line. This is the origin of the $z^4$ suppression factor. Since the $gg^{\funp\funp}$ luminosity selects $J_z=0$ states, the leading-order $gg^{\funp\funp} \to \bb$ QHC amplitude is zero. The emission of extra gluons overcomes this selection rule, but as very soft gluon emissions do not change the original structure of the $gg^{\funp\funp} \to \bb$ amplitude, it means that the cross section is suppressed by the factor $z^4$, see \cite{KRS}. 

However, the quark helicity non-conserving (QHNC) contribution, to the process of Fig.~\ref{fig:1}(d), does not contain $z^4$, but instead is suppressed by a factor $m^2_b/E_T^2$. It is given by
\be
d\hat{\sigma}^{\rm QHNC} ~=~ \frac{9}{2}~\frac{1}{(1-z)^2}~\frac{dz}{z}~\frac{dk_T^2}{k_T^2}~\frac{\alpha_S^3 m_b^2}{M^4_{\rm central}}\left(\frac{{\rm cos}^2\theta}{(1-{\rm cos}^2\theta)^2}\right)d{\rm cos}\theta,
\label{eq:HNC}
\ee 
which again allows for the emission of the $g$-jet in both the $x_1$ and $x_2$ directions. Formula (\ref{eq:HNC}) is written in the limit of soft gluon emission.  Both this contribution and the leading-order (LO)
$gg^{\funp\funp} \to \bb$ contribution are helicity non-conserving, and hence contain the $(m_b/E_T)^2$ suppression. However $\hat{\sigma}^{\rm QHNC}(\bb g)$ contains an additional ${\rm cos}^2\theta$ factor in comparison with the LO result for $gg^{\funp\funp} \to \bb$. In fact, since now the $\bb$ pair is produced in an {\it antisymmetric} colour octet state, the amplitude changes sign under the $b \leftrightarrow \bar{b}$ permutation.  That is, instead of the $1/t+1/u$ form of the LO amplitude, we now have a $1/t-1/u$ behaviour. The cos$\theta$ factor reflects the specific invariance under the rotation of the $J_z=0$ amplitude. The resulting $\sigma^{\rm QHNC}(\bb g)$ background is shown in the Table. 

Strictly speaking, this $\bb g$ contribution should be considered together with the loop correction to the exclusive $\bb$ production of Fig.~\ref{fig:1}(c), which up to now (that is at LO) represents the largest source of background. The complete virtual loop corrections to $gg^{\funp\funp} \to \bb$ is unknown, but is, at present, under study \cite{shuv}. It is anticipated that it will be negative relative to the LO amplitude, and sizeable.

Of course there are other sources of background, which were discussed in Refs. \cite{DKMOR,KRS}, for example the misidentification of gluon jets as $b$ jets. Considering the backgrounds to the exclusive signal, originating from the processes of Figs.~\ref{fig:1} and \ref{fig:2}, we find the irreducible $\bb$ background is the largest, while the contributions caused by additional soft gluons are much smaller, and do not pose a problem for experiment.

\section*{Acknowledgements}

We thank Michele Arneodo, Albert De Roeck, Brian Cox, Jeff Forshaw and Risto Orava for discussions. MGR thanks the IPPP at the University of Durham for hospitality. This work was supported by INTAS grant 05-103-7515, by grant RFBR 07-02-00023 and by the Federal Program of the Russian Ministry of Industry, Science and Technology RSGSS-5788.2006.02.


\end{document}